\documentclass[final]{aipproc}
\usepackage{bm}
\usepackage{amssymb}
\usepackage{amsmath}
\layoutstyle{8x11double}
\begin{document}

\title{Selective and optimal illumination of nano-photonic structures using optical eigenmodes}

\author{M. Mazilu}{
  address={University of St Andrews, School of Physics \& Astronomy,\\ North Haugh, St Andrews, Fife, KY16 9SS, Scotland, UK}
}
\author{K. Dholakia}{
  address={University of St Andrews, School of Physics \& Astronomy,\\ North Haugh, St Andrews, Fife, KY16 9SS, Scotland, UK}
}

\begin{abstract}
Using optical eigenmodes defined by the interaction between the electromagnetic fields and photonic structures it is possible to determine the optimal illumination of these structures with respect to a specific measurable quantity. One such quantity considered here is the electric field intensity in the hotspot regions of an array of nano-antennas.  This paper presents two possible methods, both based on optical eigenmodes, to determine the optimal and most efficient illumination that couples to a single hotspot on top of a single nano-antenna taken from an array of nano-antennas. The two methods are compared in terms of cross-talk and overall coupling efficiency. 
\end{abstract}

\maketitle

\section{Introduction}%

Structured illumination or higher order beams have found many applications in microscopy~\cite{Muller:2010p9721}, plasmonics~\cite{Gjonaj:2011dx}, metamaterials~\cite{Kao:2011vi} and more generally optical interactions~\cite{Ploschner:2011es}. Most generally, it consists in the creation of optical fields that have special properties with respect to a specific interaction. Special beam, such as the Airy beams~\cite{Baumgartl:2008kg}, Bessel beams~\cite{Mazilu:2009p9970} and Laguerre-Gaussian~\cite{MacDonald:2002p2663} beams each respond to very special needs and show particular usefulness in specific circumstances. However, the general question arises of the existence of a particular beam profile or structured illumination for each device or interaction considered. Here, we build on the optical eigenmode method to show that it is possible to define such device specific beams that are capable to selectively couple to different devices in an array of nano-photonic structures.

Intensity optical eigenmodes are electromagnetic fields that are additive in intensity when linearly superposed~\cite{Mazilu:2009p1247,Mazilu:2011p1220}. 
Here, we expand the optical eigenmode definition to treat multiple non-contiguous regions of interest over which the intensity is measured. 
In a first part, we present the general optical eigenmode method and its application to decompose target fields. In a second part, we apply the method to the numerical determination of selective structured illumination of an array of nano-antennas.

\section{Theory}

In the following we consider the structured illumination of an array of nano-antennas each capable to create an intensity hotspot in an associated region of interest $R_i$ where the index $i$ indicates the nano-antenna. To define the multi-point intensity optical eigenmode across this array, we decompose the incident electromagnetic field $\textrm{E}$ in a superposition of $N$ monochromatic ($e^{i\omega t}$) scalar ``test'' fields:
\begin{eqnarray}
{\rm E}=\sum_j a^*_j {\rm E}_j.
\label{E}
\end{eqnarray}
Here, we consider the total field intensity $\it{m}^{(I)}$ integrated over multiple regions of interest ($R_i$), as defined by:
\begin{equation}
m^{(I)}( {E})=\sum_i\int_{R_i}  {\rm E} {\rm E}^*\;d\sigma.
\label{Intensity}
\end{equation}
Each of the different region of interest $R_i$ represents the position of a hotspots that we wish to address optically. Equation (\ref{Intensity}) can be written in a general quadratic matrix form:
\begin{equation}
m^{(I)}({\rm E})=a^*_j M_{jk}a_k
\label{Intensity2}
\end{equation}
where the elements $M_{jk}$ are constructed by combining the fields ${\rm E}_j$ and ${\rm E}_k$ for $ j,k=1...N$:
\begin{equation}
M_{jk}=\sum_i\int_{R_i}  {\rm E}_j{\rm E}^*_k\; d\sigma.
\label{matrix}
\end{equation} 
The optical eigenmodes are defined by:
\begin{equation}
 {\mathbb{E}_\ell} =\frac{1}{\sqrt{\lambda^\ell}} v^*_{\ell j}{\rm E}_j 
 \label{OE2}
\end{equation}
with 
\begin{equation}
 M_{jk}{ v}_{\ell j}=\lambda^\ell { v}_{\ell k}
 \label{OE}
\end{equation}
where $\lambda^\ell$ is an eigenvalue  and ${v}_{\ell j}$ the associated eigenvector.
The matrix  $M_{jk}$ shows two important properties.

Firstly, it is Hermitian meaning that all the eigenvalues ($\lambda^\ell$) are real and can be ordered where $\lambda^{\ell=1}$ is the largest eigenvalue,  $\lambda^{\ell=2}$ the second largest eigenvalue, etc... . This means that the optical eigenmode associated with $\lambda^{\ell=1}$ describes the superposition of initial fields ${\rm E}_j$ delivering the maximal total intensity across the region of interest $R_i$. Further, considering a single region of interest would correspond to the structured illumination having the maximal coupling to the associated hotspot.

Secondly, two eigenvectors corresponding to distinct eigenvalues are orthogonal which means that: 
\begin{equation}
\sum_i \int_{R_i} {{\mathbb{E}}_j\mathbb{E}}^*_k\;d\sigma=\delta_{jk}.
\end{equation}
An unknown target field ${\rm T}$, defined in the $R_i$ regions of interest, can be decomposed onto the optical eigenmode base using its projection defined by:
\begin{equation}
c_\ell^*=\sum_i \int_{R_i}{ \rm T}\cdot {\mathbb{E}}^*_\ell \;d\sigma.
 \label{projection}
\end{equation}
where $c_\ell$ corresponds to the complex decomposition coefficients of the field ${ \rm T}$ in base $\mathbb{E}_\ell$. 
If the $\mathbb{E}_\ell$ fields form a complete base, we can perfectly reconstruct the unknown field ${\rm T}$ from the projection using ${\rm T}=c^*_\ell  {\mathbb{E}_\ell}$. We remark that the completeness of the base is dependent on the initial fields probing all the optical degrees of freedom available.  

Each of these two properties describes a possible approach to optimally couple to individual hotspots defined by the array of nano-antennas. The first property would deliver highest coupling efficiency with possible cross-talk between the different hotspots while the second property delivers the smallest possible cross talk. We remark that zero cross talk can only be achieved when the optical degrees of freedom~\cite{Piestun:2000wj} of the optical system are larger or equal then the number of hotspots. 

\section{Application}

To validate our approach, we numerically determine the electromagnetic field profile of a nano-antenna array device under coherent illumination. To achieve this, we used finite elements methods (Comsol) to simulate the reflected field of five nano-antennas illuminated by plane wave having different angle of incidence. Figure 1 shows the electric field intensity of the structure under normal illumination and we can observe the creation of a standing wave pattern on top of the nano-antennas. Here, we considered 41 different angles of incidence equally spaced and ranging from $-\pi/4$ to $\pi/4$ measured with respect to the normal. 

\begin{figure}
\includegraphics[width=.45\textwidth]{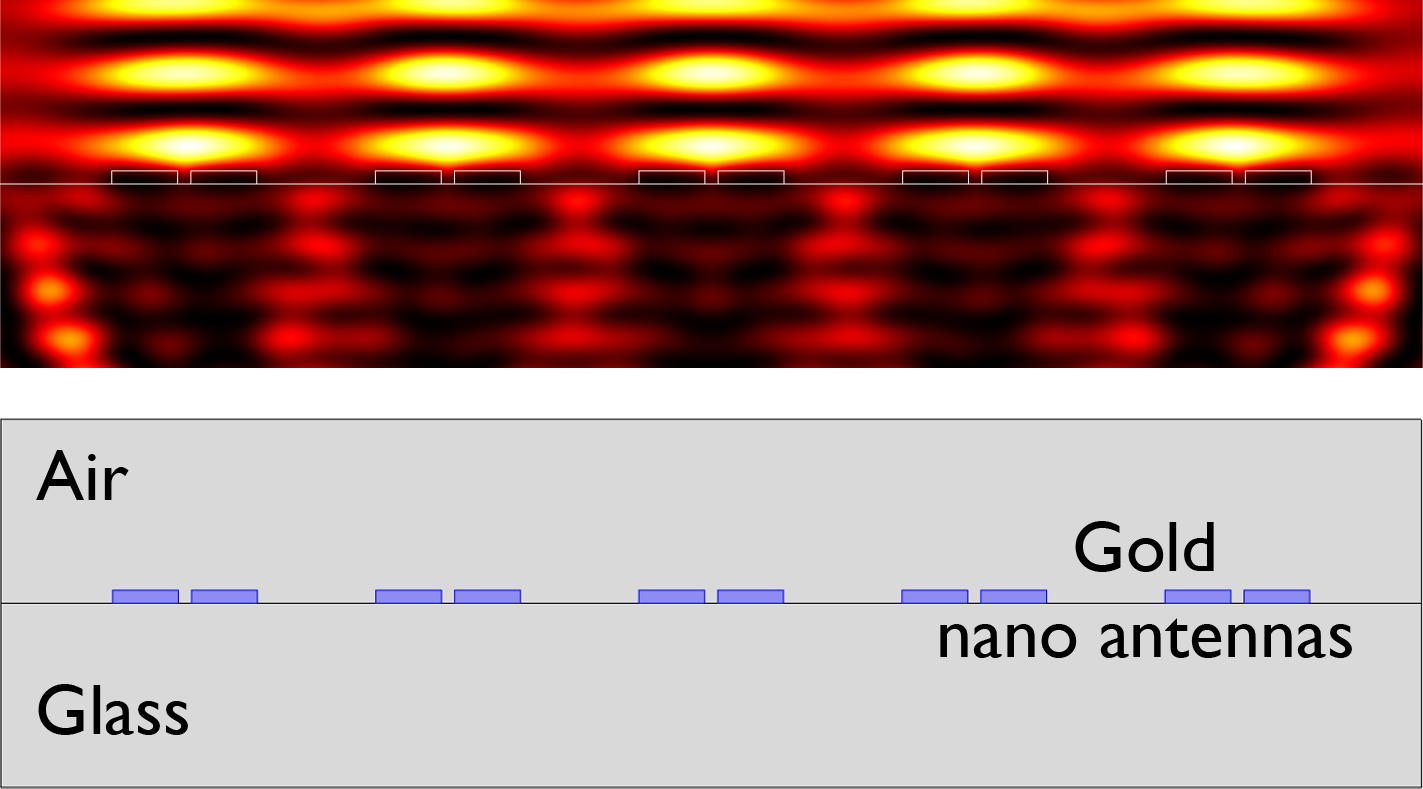}
  \caption{(top) Electric field intensity of an array of nano-antennas under plane wave illumination in normal incidence and a wavelength of $\lambda=532nm$. (bottom) Device diagram showing pairs of gold nano-antennas 50nm thick, 200nm wide with a 50nm gap. The pairs of nano-antennas are spaced 1$\mu$m apart.}
\end{figure}

\begin{figure}
\includegraphics[width=.45\textwidth]{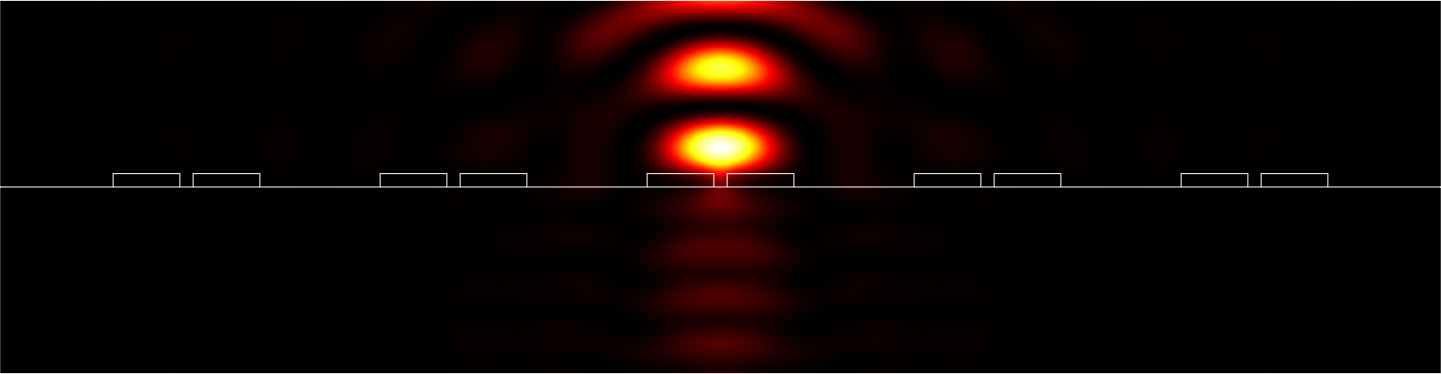}
  \caption{Electric field intensity of the single hotspot  principal eigenmode having larges eigenvalue when considering one single region of interest.}
\end{figure}

Taking into account one single region of interest corresponding to the field on top of the nano-antenna in the middle of the device, we can determine using equation (\ref{OE}) the illumination delivering the largest intensity in this region of interest. Figure 2 shows the intensity resulting from the interaction between this illumination and the nano-structure. This approach can be repeated successively with each of the different nano-antennas  determining  the illumination necessary to achieve the highest intensity in each different hotspot. 

\begin{figure}
\includegraphics[width=.45\textwidth]{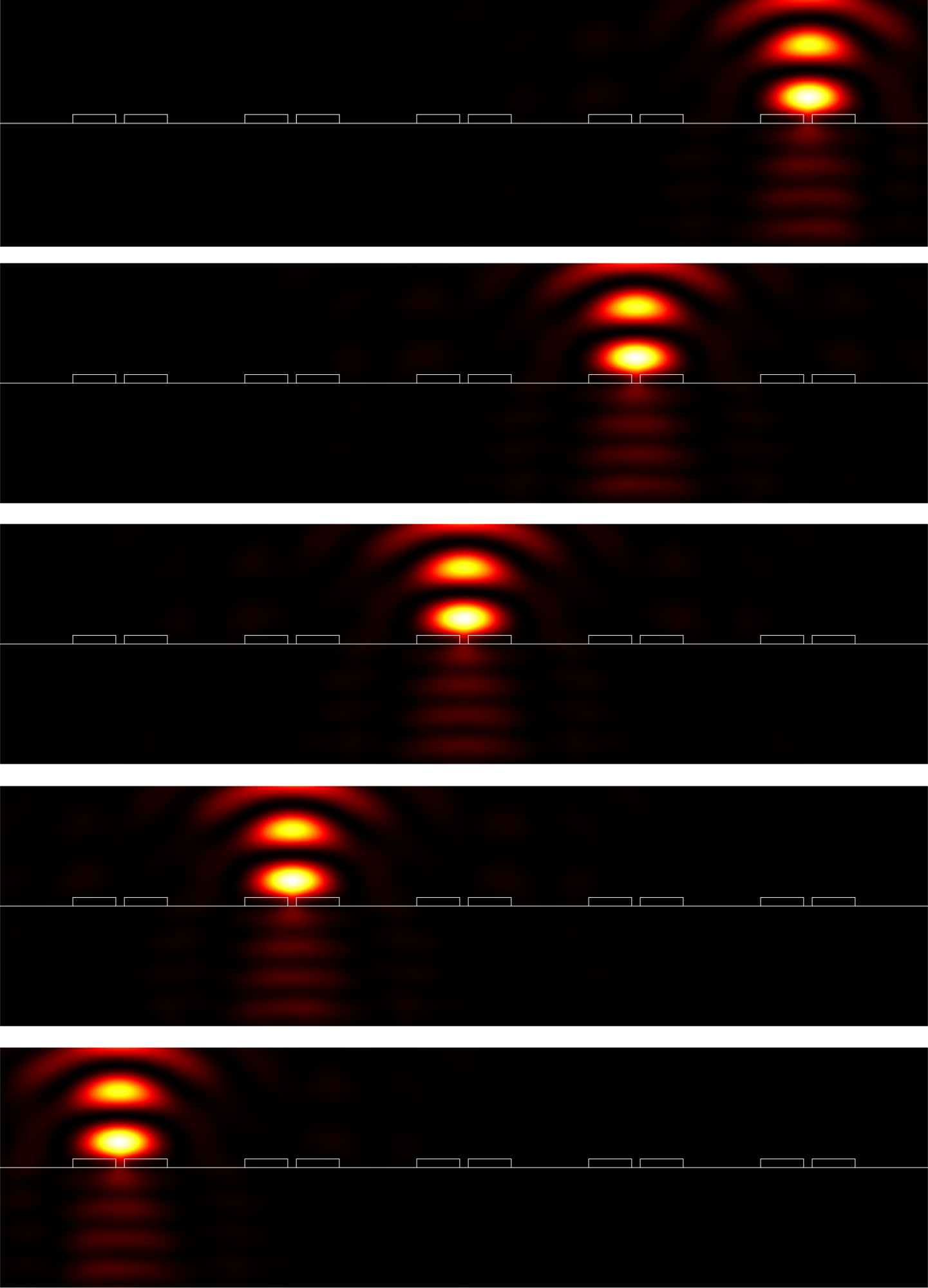}
  \caption{Electric field intensity of the normalised optical eigenmodes projected onto the target functions successively coupling to the different nano-antennas.}
\end{figure}

\begin{figure}
\includegraphics[width=.45\textwidth]{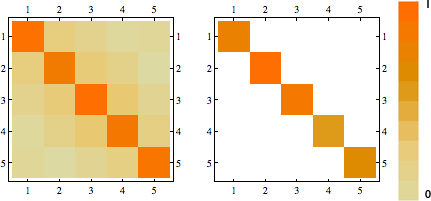}
  \caption{Cross-talk coupling efficiency for the different illuminations for (left) Figure 2 and (right) Figure 3. }
\end{figure}

The second approach is based on the orthonormal property of the optical eigenmodes as defined by equation~(\ref{OE2}). Indeed, using the projection defined by equation~(\ref{projection}) it is possible to determine the superposition of the optical eigenmodes necessary to achieve any target field. Here, we consider 5 different target fields, each one equal to one for one of the 5 nano-antennas and zero for the other 4. Figure 3 shows the field intensity achieved using these method for the 5 different cases. 

Finally, to quantify the overall coupling efficiency, we can calculate the cross-talk matrix consisting of the intensities in the different hotspots when using the different structured illuminations. Figure 4 shows this matrix for the two different approaches described here. We observe that for both methods the diagonal terms are the most important, however, the optimisation method (Figure 2) is accompanied by a few percent cross talk while the projection method (Figure 3) shows no cross talk at all. This can be understood as a question of mode purity. The projection method ensures highest possible mode purity at the expense of overall efficiency while the single hotspot eigenmode ensures highest possible coupling efficiency at the expense of mode purity. The difference between the methods becomes more pronounced when the distance between the devices decreases to smaller then the illumination wavelength. In this case, the number of optical degrees of freedom in the system decreases to the point where the projection method becomes highly inefficient while the single hotspot method induces high levels of cross-talk. 

\section{Conclusion}

We have shown two methods to determine the structured illumination of an array of nano-antennas to selectively couple to the different devices in the array. The optimisation method delivers the optical fields having the highest possible coupling efficiency to the different devices, however this approach does not eliminate cross-talk i.e. electromagnetic field in the other hotspots. The second method, is based on the projection onto the orthonormal optical eigenmodes of  target fields that are ensuring the absence of cross talk at the expense of efficiency.  

\begin{theacknowledgments}
We thank the UK Engineering and Physical Sciences Research Council (EPSRC) for funding. K.D. is a Royal Society Wolfson-Merit Award Holder.
\end{theacknowledgments}

\bibliographystyle{aipproc}   
\bibliography{tacona}

\end{document}